\documentclass[aps,prl,twocolumn,superscriptaddress,showpacs]{revtex4}
\usepackage{graphicx}
\usepackage{dcolumn}
\usepackage{bm}

\renewcommand{\v}[1]{{\bf #1}}
\newcommand{\nn}{\nonumber\\}
\newcommand{\be}{\begin{equation}}
\newcommand{\ee}{\end{equation}}
\newcommand{\ba}{\begin{eqnarray}}
\newcommand{\ea}{\end{eqnarray}}

\begin{document}

\title{Spin-polarization coupling in multiferroic transition-metal oxides}

\author{Chenglong Jia}
\email[Electronic address:$~$]{cljia@kias.re.kr}
\affiliation{School of Physics, Korea Institute for
Advanced Study, Seoul 130-012, Korea}

\author{Shigeki Onoda}
\email[Electronic address:$~$]{onoda@sss1.t.u-tokyo.ac.jp}
\affiliation{CREST, Department of Applied Physics,
University of Tokyo, Tokyo 113-8656, Japan}

\author{Naoto Nagaosa}
\email[Electronic address:$~$]{nagaosa@ap.t.u-tokyo.ac.jp}
\affiliation{CREST, Department of Applied Physics,
University of Tokyo, Tokyo 113-8656, Japan}
\affiliation{Correlation Electron Research Center, National
Institute of Advanced Industrial Science and Technology,
1-1-1, Higashi, Tsukuba 305-8562, Japan}

\author{Jung Hoon Han}
\email[Electronic address:$~$]{hanjh@skku.edu}
\affiliation{Department of Physics and Institute for Basic Science Research, \\
Sungkyunkwan University, Suwon 440-746, Korea}
\affiliation{CSCMR, Seoul National University, Seoul
151-747, Korea}
\date{\today}

\begin{abstract}
A systematic microscopic theory of magnetically induced
ferroelectricity and lattice modulation is presented for
all electron configurations of Mott-insulating
transition-metal oxides. Various mechanisms of polarization
are identified in terms of a strong-coupling perturbation
theory. Especially, the spin-orbit interaction acting on
the ligand $p$ orbitals is shown to give the ferroelectric
polarization of the spin-current form, which plays a
crucial role particularly in $e_g$ systems.
Semiquantitative agreements with the multiferroic TbMnO$_3$
are obtained. Predictions for X-ray and neutron scattering
experiments are proposed to clarify the microscopic
mechanism of the spin-polarization coupling in different
materials.
\end{abstract}


\maketitle

The coupling among the charge, spin, and orbital degrees of freedom
in Mott insulators has been one of the central issues in strongly
correlated electron systems. Despite the charge localization, its
displacement can occur accompanied by the lattice deformation as in
the dielectric or ferroelectric band insulators. In the case of Mott
insulators, the charge displacement is usually triggered by a
non-trivial low-energy spin and/or orbital structure, which is
unveiled within an energy scale of the exchange interaction $J\sim
V^4/\Delta^3$ with the hybridization $V$ and the charge transfer
energy $\Delta$ between the transition-metal (TM) $d$ and ligand (L)
$p$ orbitals. There, the ferroelectricity and the magnetism
sometimes appear concomitantly as a spontaneous magnetoelectric
effect dubbed multiferroic order~\cite{curie,fiebig,tokura}.
Usually, the largest magnetoelectric force is given by the
exchange-striction. Spatial imbalance in the symmetric magnetic
exchange leads to a lattice modulation. When it breaks the inversion
symmetry, a finite net dipole moments appears as the ferroelectric
polarization.

Recent observations of the ferroelectric polarization at
the onset of the spiral magnetic order in
TbMnO$_3$~\cite{kimura,kenzelman,arima} and
Ni$_3$V$_2$O$_8$~\cite{lawes} have attracted revived
interests in multiferroic behavior from the viewpoints of
the fundamental physics as well as
applications~\cite{tokura}. Microscopic~\cite{KNB} and
phenomenological~\cite{mostovoy,harris} theories have
revealed that spiral and conical spin structures can couple
to the polarization $\bm{P}$ in the absence of any symmetry
breaking through $\bm{P}\cdot\bm{J}_s$, where
$\bm{J}_s\propto\bm{e}\times(\bm{S}_{\bm{r}}\times\bm{S}_{\bm{r}+\bm{e}})
\propto\bm{S}(\bm{\nabla}\cdot\bm{S})-(\bm{S}\cdot\bm{\nabla})\bm{S}$
is the spin current produced between the two spins
$\bm{S}_{\bm{r}}$ and $\bm{S}_{\bm{r}+\bm{e}}$ at sites
$\bm{r}$ and $\bm{r+e}$, respectively. This coupling has
been derived by taking account of the spin-orbit coupling
$\lambda$ for a model of the $t_{2g}$ orbitals of TM ions
hybridized with the L $p$ orbitals~\cite{KNB,JONH}, an
$e_g$ system coupled to the localized $t_{2g}$
spin~\cite{Hu}, and a $sp$ hybridized model~\cite{kaplan}.
Effects of the exchange-striction and the
Dzyaloshinskii-Moriya (DM) interaction have also been
discussed in the context of the multiferroic
$R$MnO$_3$~\cite{dagotto}.

In view of the recent proliferation in the number of
multiferroic materials, a proper classification scheme
seems to be in order. In this Letter, we propose a
systematic microscopic theory of the magnetically induced
polarization and lattice modulation in multiferroic
transition-metal oxides, by means of a strong-coupling
expansion in both $V/\Delta$ and $\lambda/\Delta$. We
classify various microscopic mechanisms for the
polarization and their distinct features. The
classification scheme offers a way of identifying the
material-specific mechanism by probing the magnetically
induced non-uniform shifts of the L ions. As the specific
demonstration, key experimental observations on the
polarization and the lattice modulation in TbMnO$_3$ are
analyzed.

\begin{figure}[t]
\begin{center}
\includegraphics[width=8cm]{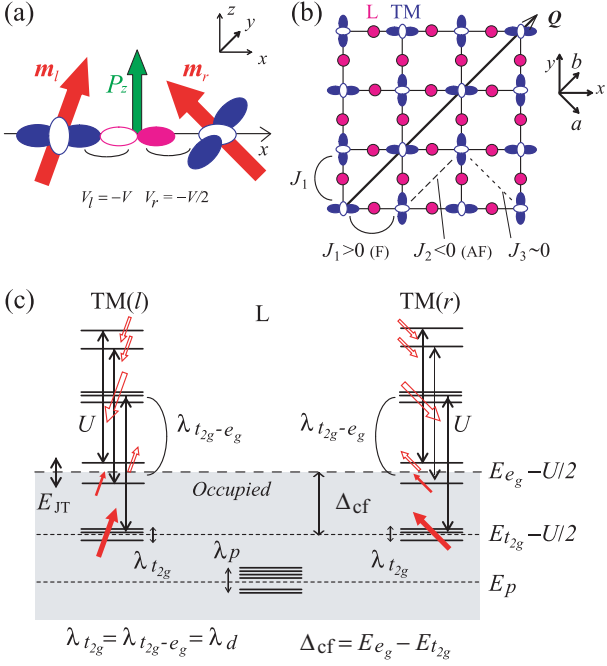}
\end{center}
\caption{(color online) (a) The TM-L-TM cluster model with the rod-type $d_{3x^2-r^2}$/$d_{3y^2-r^2}$ staggered orbital order under a noncollinear spin configuration with the associated electric polarization.
(b) The lattice structure of the perovskite system within the $xy$ (or $ab$) plane. The staggered orbital order for TbMnO$_3$ is also shown.
(c) The level scheme for the $t_{2g}^3e_g^1$ high-spin ($S=2$) configuration. }
\label{fig:Model}
\end{figure}

Let us consider a cluster composed of TM ions at $\bm{r}$
and $\bm{r}+\bm{e}$ hybridized through a L ion at the
center $\bm{r}+\bm{e}/2$ as shown in Fig.~\ref{fig:Model}
(a). For simplicity, we neglect a deviation of the bond
angle away from 180$^\circ$. The dipole moment
$\bm{P}_{\bm{r}+\bm{e}/2}$ induced at the ligand site
$\bm{r}+\bm{e}/2$ is generally expressed as

\begin{eqnarray}
\bm{P}_{\bm{r}+\bm{e}/2} &=& \left[
P^{\mathrm{nm}}_{\bm{r},\bm{r}+\bm{e}}+P^{\mathrm{ms}}_{\bm{r},\bm{r}+\bm{e}}
(\bm{m}_{\bm{r}}\cdot\bm{m}_{\bm{r}+\bm{e}})\right]\bm{e}\nn
&&+P^{\mathrm{orb}}_{\bm{r},\bm{r}+\bm{e}}
\left[(\bm{e}\cdot\bm{m}_{\bm{r}})\bm{m}_{\bm{r}}-
(\bm{e}\cdot\bm{m}_{\bm{r}+\bm{e}})
\bm{m}_{\bm{r}+\bm{e}}\right] \nn
&&+P^{\mathrm{sp}}_{\bm{r},\bm{r}+\bm{e}}
\bm{e}\times(\bm{m}_{\bm{r}}\times\bm{m}_{\bm{r}+\bm{e}})
\label{eq:P}
\end{eqnarray}
up to second order in the local spin moments
$\bm{m}_{\bm{r}}$ and $\bm{m}_{\bm{r}+\bm{e}}$.
Here, $P^{\mathrm{nm}}\propto V/\Delta$ is the nonmagnetic
term that occurs regardless of the magnetic properties.
$P^{\mathrm{ms}}\propto(V/\Delta)^3$ represents the
magnetostriction term. These effects do not demand the
spin-orbit interaction. The orbital term
$P^{\mathrm{orb}}\sim \mathrm{min}(\lambda/V, 1)
(V/\Delta)$ has been obtained in the study of partially
filled degenerate $t_{2g}$ orbitals\cite{JONH}. Finally,
the spin-current term
$P^{\mathrm{sp}}\propto(\lambda/\Delta)(V/\Delta)^3$ was
demonstrated for the degenerate $t_{2g}$
model~\cite{KNB,JONH} and, as will be shown below, exists
for any Mott-insulating electron configuration.
$P^{\mathrm{sp}}$ and $P^{\mathrm{orb}}$ are \textit{even}
under the spatial inversion and found only in the presence
of spin-orbit interaction, while $P^{\mathrm{nm}}$ and
$P^{\mathrm{ms}}$ are \textit{odd} and present only in the
absence of the inversion symmetry.

On an array of $\cdots-$TM$-$L$-$TM$-$L$-\cdots$ with (at
most) two inequivalent L positions due to the orthorhombic
distortion and/or staggered orbital order, the various
mechanisms of polarization are manifested as Fourier
harmonics in X-ray or neutron scattering probes. Assuming a
general conical structure $\bm{m}_{\bm{r}}
=\bm{m}_0+\bm{m}_1\cos\bm{Q}\cdot\bm{r}-\bm{m}_2\sin\bm{Q}\cdot\bm{r}$,
we obtain the nontrivial Fourier components
$\bm{P}(\bm{q})=\sum_{\bm{r},\bm{e}}\bm{P}_{\bm{r}+
\bm{e}/2}e^{-i\bm{q}\cdot(\bm{r}+\bm{e}/2)}$ shown in
Table~\ref{table:mechanism}, with $\bm{m}_\pm=\bm{m}_1\pm
i\bm{m}_2$ and $\bm{P}_{\bm{r}+ \bm{e}/2}$ given in Eq.
(\ref{eq:P}). To compensate the electrostatic energy loss,
each L ion tends to shift towards the induced dipole moment
and cause the lattice modulation with the same wave
vectors. $P^\mathrm{ms}$ gives a collinear lattice
modulation along the cluster direction with wave vectors
$\pi \bm{e} + \bm{Q}$ and $\pi\bm{e}+2\bm{Q}$, where the
shift of $\pi \bm{e}$ is due to the present of two
equivalent $L$ sites. It becomes uniform for the special
case $\bm{Q}\cdot \bm{e}=\pi/2$, or period four for
magnetism. This case is relevant for HoMnO$_3$ as pointed
out by Sergienko \textit{et al.} recently\cite{dagotto}.
Under the spiral (conical) magnetic structure,
$P^\mathrm{orb}$ yields a spiral, even noncoplanar,
modulation, while $P^\mathrm{sp}$ gives a uniform (conical)
modulation. $P^\mathrm{orb}$ is found only in the
unfulfilled $t_{2g}$ system\cite{JONH}. These results can
be used to classify the mechanisms of the magnetoelectric
couplings using X-ray and neutron scattering experiments.

Equation~(\ref{eq:P}) can be derived from the TM-L-TM
cluster model\cite{KNB,JONH} composed of the L $p$ orbitals
and the two TM $d$ orbitals which are hybridized with L
through the Slater-Koster parameters $V_{pd\sigma}$ and
$V_{pd\pi}$. We assume an effective Zeeman field
$(U/2)\bm{\hat{m}}_a$ ($a=l/r$ for the left/right TM ion)
with an energy separation $U$, which originates from the
local Coulomb repulsion and the Hund coupling in the
magnetically ordered phase. The local energy levels of the
TM $d_\alpha$ and $L$ $p$ orbitals are $E_{d_\alpha}$ and
$E_p$, respectively, while the spin-orbit couplings for $d$
and $p$ orbitals are denoted $\lambda_d$ and $\lambda_p$.

\begin{table}
  \begin{center}
    \begin{tabular}{c|c|c}
      \hline\hline
      $\bm{P}(\bm{q})$ & Mechanisms & Polarization\\
      \hline
      $\bm{P}(\bm{0})$ & $P^\mathrm{sp}$ & $-\bm{e}\times(\bm{m}_1\times\bm{m}_2)\sin\bm{Q}\cdot\bm{e}$\\
      \hline
      $\bm{P}(\pm(\pi\bm{e}+\bm{Q}))$ & $P^\mathrm{ms}$ & $\bm{e}\left[\bm{m}_0
      \cdot(\bm{m}_1\pm i\bm{m}_2)\right]\cos [\bm{Q}\cdot\bm{e} /2]$\\
      \hline
      $\bm{P}(\pm(\pi\bm{e}+2\bm{Q}))$ & $P^\mathrm{ms}$ & $\bm{e}(\bm{m}_1^2+\bm{m}_2^2)/4$\\
      \hline
      $\bm{P}(\pm\bm{Q})$ & $P^\mathrm{orb}$ & $\mp i\left[(\bm{e}\cdot\bm{m}_0)\bm{m}_\pm+(\bm{e}\cdot\bm{m}_\pm)\bm{m}_0\right]$\\
      \cline{2-3}
      & $P^\mathrm{sp}$ & $\mp\frac{i}{2}\bm{e}\times\left(\bm{m}_0\times\bm{m}_\mp\right)
      \sin[\bm{Q}\cdot\bm{e}/2]$\\
      \hline
      $\bm{P}(\pm2\bm{Q})$ & $P^\mathrm{orb}$ & $\mp\frac{i}{2}\left[
      (\bm{e}\cdot\bm{m}_\pm)\bm{m}_\pm\right]\sin\bm{Q}\cdot\bm{e}$\\
      \hline\hline
    \end{tabular}
  \end{center}
\caption{Wave vectors and directions of the lattice
modulation originating from the mechanisms $P^\mathrm{ms}$,
$P^\mathrm{orb}$, and $P^\mathrm{sp}$. $P^\mathrm{nm}$
always gives a modulation with wave vector $\pi \bm{e}$.}
\label{table:mechanism}
\end{table}

The mechanisms responsible for the spin-induced dipole
moments through spin-orbit coupling can be classified as in
Table \ref{table:source}. Spin-orbit interaction
$\lambda_{t_{2g}}$ within the degenerate $t_{2g}$ manifold
can be a source of polarization, $P^\mathrm{sp}$ and
$P^\mathrm{orb}$\cite{KNB,JONH}, when the $t_{2g}$ orbital
degrees of freedom are unquenched, e.g. $t_{2g}^n$ with
$n=1,2,4,5$. It was derived in a previous paper~\cite{JONH}
that $P^\mathrm{orb}$ and $P^\mathrm{sp}$ appear in
proportion to
$\mathrm{min}(\lambda_d/V_{pd\pi},1)(V_{pd\pi}/\Delta)$ and
$(\lambda_d/\Delta)(V_{pd\pi}/\Delta)^3$, respectively, in
the $t_{2g}^5$ configuration, where $\Delta = E_{t_{2g}} +
U/2 - E_p$. The $t_{2g}^2$ configuration is related to
$t_{2g}^5$ with the change in the charge transfer energy to
$\Delta = E_{t_{2g}} - U/2 - E_p$, and the $t_{2g}^1$ to
$t_{2g}^4$ in the same manner. Another symmetry operation
can be used to show that $P^\mathrm{sp}$ and
$P^\mathrm{orb}$ for $t_{2g}^4$ is related to those of
$t_{2g}^5$ by a sign change, as well as those for
$t_{2g}^1$ to $t_{2g}^2$. Due to the strong Hund coupling,
the $t_{2g}^3$ and $t_{2g}^6$ configurations represent the
orbital quenched case with no polarization arising from
orbital deformation.

The spin-orbit interaction within the $e_g$ manifold
vanishes and has no impact on spin-polarization coupling.
In this case, the spin-orbit interaction $\lambda_p$ at the
ligand oxygen site can mediate the spin-polarization
coupling $P^\mathrm{sp}$, as we demonstrate below. In
addition, mixing of an occupied $e_g$ level and an
unoccupied $t_{2g}$, or of an unoccupied $e_g$ and an
occupied $t_{2g}$, can occur due to the spin-orbit
interaction $\lambda_{t_{2g}-e_g}$ in the given TM ion and
contribute to the polarization $P^\mathrm{sp}$. Such
mechanism was demonstrated for TbMnO$_3$ recently\cite{Hu}.
Both mechanisms described in this paragraph should apply
for arbitrary $d^n$ configuration, as long as the net
magnetic moment remains nonzero to guarantee magnetic
ordering.

\begin{table}
\begin{center}
\begin{tabular}{ccc}
\hline\hline
 Spin-orbit & Configuration & Polarization\\
\hline $\lambda_{t_{2g}}$ & $t_{2g}^n$ ($n=1,2,4,5$) &
$P^\mathrm{sp}$,
$P^\mathrm{orb}$\\
\hline
$\lambda_{t_{2g}-e_g}$ & arbitrary $d^n$, nonzero $S$ & $P^\mathrm{sp}$ \\
\hline
$\lambda_p$ & arbitrary $d^n$, nonzero $S$ & $P^\mathrm{sp}$\\
\hline\hline
\end{tabular}
\end{center}
\caption{Classification of the spin-orbit interactions and
the corresponding types of induced polarization and the
relevant $d$ electron configuration. Nonzero net magnetic
moment for the $d$ electrons, $S \neq 0$, is required to
generate magnetic ground states. $\lambda_{t_{2g}-e_g}$
($\lambda_p$) effects decrease (increase) as the electron
number $n$ increases.}\label{table:source}
\end{table}

A low-$n$ transition metal involving Ti or V will be the
likely candidate to observe $t_{2g}$-mediated
$P^\mathrm{sp}$ and $P^\mathrm{orb}$. A spin-canted phase
in YVO$_3$ was found recently\cite{YVO}. For such a state
we argue that $P^\mathrm{orb}$, which is much larger than
$P^\mathrm{sp}$ on the atomic scale, is detectable in the
microscopic X-ray or neutron probes. The specific
dependence of $P^\mathrm{orb}$ on the local magnetic
structure given in Eq. (\ref{eq:P}) and in Table
\ref{table:mechanism} can be checked experimentally.

Recently, a number of $e_g$ systems with filled $t_{2g}$
came to be classified as a multiferroic: a nickelate
Ni$_3$V$_2$O$_8$~\cite{lawes} with $d^8$, and a cuprate
LiCuVO$_4$~\cite{LiCuVO4} with $d^9$. We propose that the
oxygen spin-orbit-mediated mechanism should play an
important role for these materials. For instance, the $d^9$
Cu sites can be modeled as $d_{x^2-y^2}$ orbitals at the TM
sites in the cluster model. Taking $-V_l = V_r =
V_{pd\sigma}$\cite{apology}, $P^\mathrm{sp}$ is given
by\cite{elsewhere}

\begin{equation}
P_{\lambda_p}^\mathrm{sp} =- 2\sqrt{2}(L_{r,z} + L_{l,z})
\frac{\lambda_p}{\Delta}\left(\frac{V_{pd\sigma}}{\Delta}\right)^3
\label{eq:d9:P^sp}
\end{equation}
where $\Delta = E_{e_g}\!+\!U/2\!-\!E_p$, and $E_{e_g}$ is
the energy of the relevant $e_g$ orbital. Here $L_{a,z}$
denotes the integral $L_{a,z} = \langle
d_{a}|z|p_{z}\rangle$ with $a=r,l$. The $d^8$ contains two
$e_g$ orbitals with $S=1$, that can be decomposed as
$d_{3x^2 -r^2}$ and $d_{y^2-z^2}$ in the degenerate case.
Of these, the latter does not couple to the oxygen orbitals
and one is left with a single $d_{3x^2-r^2}$ orbital per
site with the polarization again given by Eq.
(\ref{eq:d9:P^sp}). No $P^\mathrm{orb}$ term is found for
pure $e_g$ models with oxygen spin-orbit interaction.

The $d^7$ case with a single $e_g$ electron is subject to
the Jahn-Teller distortion and the lifting of the orbital
degeneracy. When the same $e_g$ orbitals are occupied for
the adjacent TM sites, one again has Eq.
(\ref{eq:d9:P^sp}). If not, the appropriate cluster model
should involve different orbitals for the right and left
TM's, and the inversion symmetry breaking yields non-zero
values of

\be P^\mathrm{nm} \!=\! (L_{r,x} \! -\!
L_{l,x})\frac{V_{pd\sigma}}{\Delta}, ~~ P^\mathrm{ms}
\!=\!{1\over 4} (L_{l,x} \!-\! L_{r,x})
\Bigl(\frac{V_{pd\sigma}}{\Delta}\Bigr)^3,
\label{eq:d9:P^ms} \ee
with $L_{a,x} =\langle d_a | x | p_x \rangle$. On the local
scale both terms, being independent of spin-orbit
interaction, are greater than $P_{\lambda_p}^\mathrm{sp}$.
Except for special circumstances, however, $P^\mathrm{nm}$
and $P^\mathrm{ms}$ are finite-$\bm{q}$ modulations with
zero macroscopic average. Even for the $d^9$ and $d^8$
cases, the orthorhombic distortion displaces the oxygen
position away from the middle of the TM ions, then one has
the appearance of Eq. (\ref{eq:d9:P^ms}) in the
polarization.

As a concrete example of the $\lambda_{t_{2g}-e_g}$
mechanism at work, we discuss the case of $t^3_{2g}e_g^1$
configuration found in TbMnO$_3$ with Mn$^{3+}$ ion.
Because of the large Jahn-Teller effect, TbMnO$_3$ exhibits
the rod-type $d_{3x^2-r^2}$/$d_{3y^2-r^2}$ orbital ordering
shown in Fig.~\ref{fig:Model}(b) far above the room
temperature\cite{OO}. Accordingly we assume that the
$d_{3x^2-r^2}$ and $d_{3y^2-r^2}$ orbitals are occupied at
the $l$ and $r$ sites of the TM-L-TM cluster. Assuming $V_l
= V_r = -V_{pd\sigma}$, one obtains polarization of the
spin-current form\cite{elsewhere}

\ba P_{t_{2g}\!-\!e_g }^\mathrm{sp} =  - \sqrt{3}
L^{\prime}_{z} {\lambda_d  \over  U \!- \!
\Delta_{cf}\!+\!E_\mathrm{JT}/2 }\left( {V_{pd \sigma }
\over \Delta } \right)^3 \label{eq:t2gegPsp}\ea
where $\Delta_{cf}$ is the crystal field gap
$E_{e_g}-E_{t_{2g}}$, $E_\mathrm{JT}$ is the Jahn-Teller
energy splitting, and $L^{\prime}_{z} = \langle d_{l,zx} |
z | p_x \rangle =\langle d_{r,zx} | z | p_x \rangle
$\cite{Hu,elsewhere}. A similar term is obtained for
uniform orbital order, i.e. same $d$-orbitals for both TM
ions in the cluster. No orbital polarization
$P^\mathrm{orb}$ is induced from the
mixing\cite{elsewhere}.

In the rest of the paper, we compare the experimental
situation with the theoretical analysis for TbMnO$_3$,
which probably presents the most complex situation with
orbital ordering, unfulfilled $t_{2g}$ and $e_g$ levels,
and $t_{2g}\!-\!e_g$ mixing. The two terms, Eq.
(\ref{eq:t2gegPsp}) and Eq. (\ref{eq:d9:P^sp}), combine
constructively to produce $P^\mathrm{sp}$, the uniform
polarization. Quantitative agreement with the measured
uniform polarization value in TbMnO$_3$ is obtained as
follows.

Matrix elements needed to evaluate Eq. (\ref{eq:d9:P^sp})
and Eq. (\ref{eq:t2gegPsp}) are calculated by taking the
TM-O bond length $1.9\AA$ and the Clementi-Raimondi
effective charges~\cite{Z}
$Z_{\mathrm{Mn}3d}^{\mathrm{eff}}=10.53$ and
$Z_{\mathrm{O}2p}^{\mathrm{eff}}=4.45$. Other parameters
are chosen as $V_l =2V_r = -V_{pd\sigma}=-1.2$ eV, $U=3$ eV
and $E_{e_g}-E_p = 2$ eV. The spin-orbit coupling for the
oxygen $2p$ and TM $3d$ orbitals are chosen from
$\lambda_{L} \equiv 2 \Delta E_L / (2L +1)$, where $L =1,2$
for $p$ and $d$ orbitals. The spin-orbit energy splitting
is given by $\Delta E_p =$ 37 meV ($\lambda_p =$ 25 meV),
and $\Delta E_d =$ 120 meV ($\lambda_d = $ 48 meV),
respectively\cite{Lambda}. With these values we obtain the
uniform polarization along the $c$ axis
$P^\mathrm{sp}_{\lambda_p} \sim 130$ $\mu$C/m$^{2}$ from
the oxygen spin-orbit interaction, and
$P^\mathrm{sp}_{t_{2g}\!-\!e_g} \sim 860$ $\mu$C/m$^2$ from
the $t_{2g}\!-\!e_g$ mixing with $\Delta_{cf} = 2$ eV and
$E_\mathrm{JT} = 1$ eV. The net value $P^\mathrm{sp}\sim$
990 $\mu$C/m$^2$ multiplied with $|\bm{m}_{\bm{r}}\times
\bm{m}_{\bm{r}+\bm{e}}| \approx \sin (0.28\pi)$ in Eq.
(\ref{eq:P}) gives the uniform $c$-axis polarization of
$\sim$ 760 $\mu$C/m$^2$, in reasonable agreement with the
experimental value in TbMnO$_{3}$\cite{kimura,kenzelman}
($\sim$700 $\mu$C/m$^2$ at 10K). The direction of the
polarization is also consistent with the experiment.

\begin{figure}[b]
\begin{center}
\includegraphics[width=8.5cm]{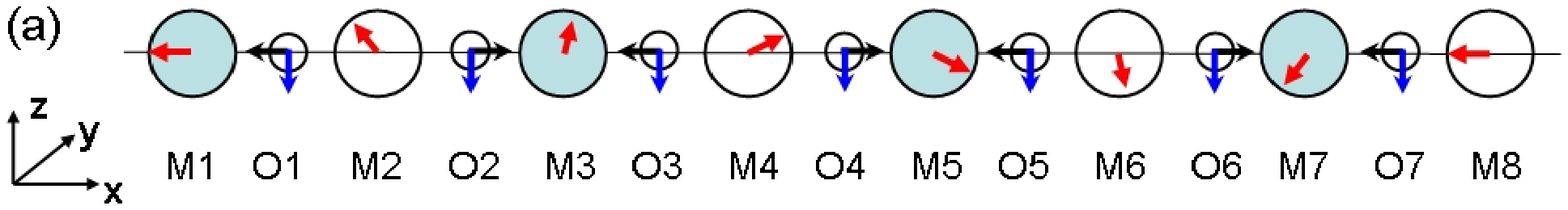} \\[10pt]
\includegraphics[width=8.5cm]{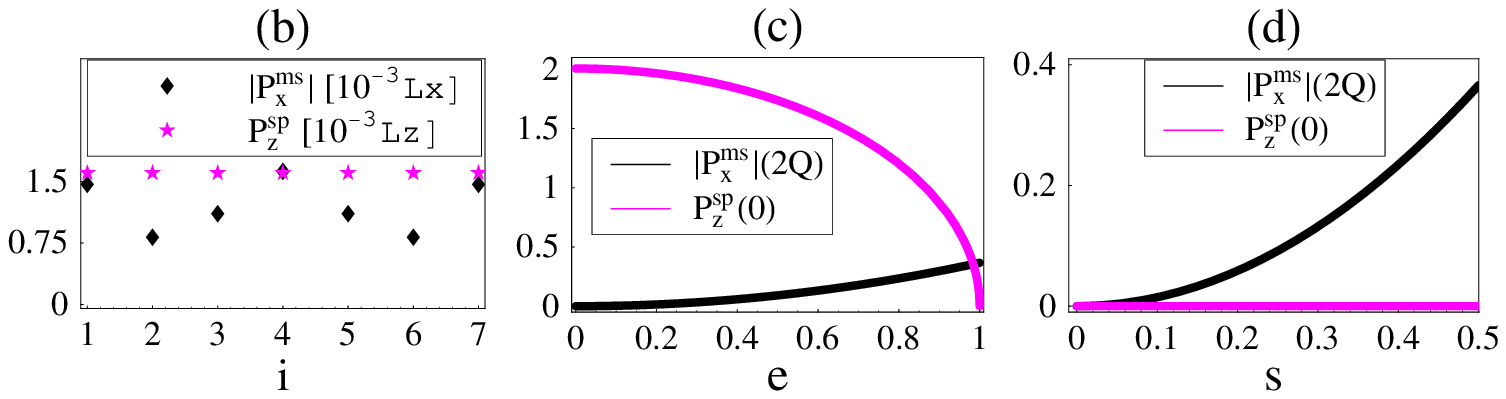}
\end{center}
\caption{(color online) (a) Elliptic spiral spins (red
arrows) with the ellipticity $e=0.6$ and the associated
local dipole moments arising from oxygen spin-orbit
interaction are shown as $P^\mathrm{sp}_z$ (blue arrows,
enlarged by $10^{3}$ times) and $P^\mathrm{ms}_x$ (black
arrows) for the observed spin modulation period ($Q\approx
0.28 \pi$) as in TbMnO$_3$ at low temperature. Two types of
Mn sites exists (blue and white) due to orbital ordering.
Statistical weights are introduced to account for elliptic
spins. $P_x^\mathrm{ms}$ alternates its direction due to
staggered orbital order. (b) Induced dipole moments
$P_{z,i}^{\mathrm{sp}}$ and $| P_{x,i}^{\mathrm{ms}} |$ at
the oxygen site $O_i$ in TM$_{i}$-O$_i$-TM$_{i+1}$ cluster
in units of $10^{-3}L_{z}$ and $10^{-3}L_{x}$,
respectively, and $L_{z(x)}$ is defined by $L_{l,z(x)}$.
(c,d) Main Fourier components of the dipole moments, at
$q=2Q$ for $|P_{x,i}^\mathrm{ms}|$ and at $q=0$ for
$P_{z,i}^\mathrm{sp}$, as functions of $e$ for $T < T_{L}$
in (c) and $S$ with $e=1$ for $T_{L} < T < T_{N}$ in (d).
Note that $e$ in (c) and $1/2 - S$ in (d) decrease with
lowering temperature. This figure is worked out for the
orbital ordered model of TbMnO$_3$ (Figs.~\ref{fig:Model}) with $(U, \Delta_{cf},E_{JT},V_{pd\sigma}, \lambda_p,
\lambda_d ) = (3.0,2.0,1,1.2,0.025,0.048)$ eV. }
\label{fig:Elliptic}
\end{figure}

The X-ray intensity of the $2\bm{Q}$ lattice modulation in
TbMnO$_3$ increases monotonically between the Neel
temperature $T_N$ for the collinear magnetic phase, and
$T_L$, the latter signaling the onset of ferroelectricity,
as the temperature is lowered. Then the signal decreases
with lower temperature in the non-collinear, ferrroelectric
phase, $T<T_L$. Both features can be understood by taking
the view that the non-uniform component arises from the
magnetostriction term, due to the breaking of the inversion
symmetry in each TM-L-TM cluster from the orbital ordering
and the orthorhombic distortion.

The spin configuration in the magnetic phase $T<T_N$ can be
described by the ordered spin moment, $\langle \v s_i
\rangle = S(\hat{b} \cos \theta_i + \hat{c} \sqrt{1-e^2}
\sin \theta_i$). It evolves from collinear ($e=1$) to
nearly circular ($e=0$) pattern with decreasing
temperature~\cite{arima}. In the collinear phase, one has
$e=1$ while the magnitude $S$ increases monotonically at
the lower temperature. The TM-L-TM cluster calculation can
be generalized to handle these cases, both collinear and
elliptic, by writing a self-consistent equation $\langle \v
s_i\rangle={\int d\Omega_i \v m_i w_i \over \int d\Omega_i
w_i}$ with a weight $w_i=\exp[ \beta_i \langle\v s_i\rangle
\cdot \v m_i ]$. Average over the O(3) vector $\v m_i$
determines the appropriate $\beta_i$. The dipole moment in
the collinear/elliptic spin configuration can be obtained
by taking the average of Eq. (\ref{eq:P}) over the spins
$\bm{m}_{\bm{r}}$ and $\bm{m}_{\bm{r}+\bm{e}}$ with weights
$w_{\bm{r}}$ and $w_{\bm{r}+\bm{e}}$, for the magnetic
period $\sim 0.28\pi$ configuration of the low-temperature
TbMnO$_3$.

As shown in Fig. \ref{fig:Elliptic} (d), in the collinear phase,
($e=1$), the $2\bm{Q}$ intensity arising from the magnetostriction
$P^{\mathrm{ms}}$ monotonously increases with the ordered spin
amplitude $S$. Then in the non-collinear phase ($e<1$), the
$2\bm{Q}$ intensity arising from $P^\mathrm{ms}$ decreases for
smaller $e$ corresponding to a lower temperature (Fig.
\ref{fig:Elliptic} (c)). The direction of the lattice modulation in
TbMnO$_3$ has not been clearly determined yet~\cite{Arima}. A proper
identification of the lattice modulation direction in the future can
be used to discriminate the different scenarios of magnetoelectric
coupling.

The authors acknowledge fruitful discussion with T. Arima, Y.
Tokura, F. Ishii, and T. Ozaki. H. J. H. was supported by the Korea
Research Foundation through Grant No. KRF-2005-070-C00044. The work
was partly supported by the Grant-in-Aids from under the Grant No.
15104006, 16076205, and 17105002, and NAREGI Nanoscience Project
from the Ministry of Education, Culture, Sports, Science, and
Technology, Japan.



\begin{thebibliography}{99}
\bibitem{apology} For $V_l =V_r = - V_{pd\sigma}$ one has to replace $L_{r,z(x)}$
  with its negative value. Generic expressions for arbitrary $V_l$ and $V_r$
  are not available.
\end{thebibliography}
\end{document}